\documentstyle[12pt,epsfig]{article}
\begin{document}
\baselineskip=0.6cm
\begin{flushright}
Accepted for publication by Physical Review D.\\
Revised on March 10, 2000.  \vspace{5mm}\\
\end{flushright}
\vspace{.5in}
\begin{center}
{\Huge Properties of Regge Trajectories }\\
\vspace{.25in}
{\Large Alfred Tang and John W. Norbury}\\
\vspace{.15in}
{\em Physics Department, University of Wisconsin - Milwaukee,}\\
{\em P. O. Box 413, Milwaukee, WI 53201.}\\
Emails: atang@uwm.edu, norbury@uwm.edu
\end{center}
\vspace{5mm} \noindent
\begin{center}
{\bf Abstract}
\end{center}
\noindent
Early Chew-Frautschi plots show that meson and baryon Regge
trajectories are approximately linear and non-intersecting.  In this paper,
we reconstruct all Regge trajectories from the most recent data.  Our plots
show that meson trajectories are non-linear and intersecting.  We also show
that all current meson Regge trajectories models are ruled out by data.

\vspace{5mm} \noindent

\section{Introduction}

The topic of Regge trajectories was an active area of research in the 1960's.
But there is a resurgence of interest in Regge trajectories because of the
quantity of new data that need analysis and the new quark models need more
complete experimental fits for testing.  Despite these recent interests, some 
authors~\cite{salvo94,andreev99} are still using old data to construct 
Chew-Frautschi plots.  In this paper,
we reconstruct all Regge trajectories with the most recent data and elucidate
the principles of their construction.  At the end, we explain why all current
meson Regge trajectories models are ruled out by data.

\begin{center}
\bf 1.1 Theoretical Developments
\end{center}

This paper is concerned with the properties of Regge trajectories which
are graphs of the total quantum number 
$J$ versus mass squared $M^{2}$ over a set of particles of fixed principal 
quantum number $N$, isospin $I$, dimensionality of the symmetry
group $D$, spin-parity and flavor. 
Variations in $J$ and $L$ $(J=L+S)$ are equivalent when $S$ is fixed.  

Scattering processes are usually analyzed by the method of 
partial-waves~\cite{chew61,sakurai85}.  The wavefunction in 
the far zone has the form
\begin{equation}
\psi ({\bf r}) \simeq e^{i{\bf k}\cdot{\bf r}} +
f(k,\cos\theta) \frac{e^{i{\bf k}\cdot{\bf r}}}{r},
\end{equation}
where $\theta$ is the angle between the wave vector $\bf k$ and the position
vector $\bf r$.  In the case of bound states, the plane wave term is absent.
The form factor $f$ is written as a sum of partial-waves as
\begin{eqnarray}
f(k^{2},\cos\theta)&=&\sum^{\infty}_{l=0} (2l+1)a_{l}(k^{2})P_{l}(\cos\theta), 
\label{f} \\
a_{l}(k^{2})&=&{1\over 2}\int^{1}_{-1} f(k^{2},\cos\theta)
P_{l}(\cos\theta)d\cos\theta,
\end{eqnarray}
where $l$ is the orbital angular momentum quantum number and $P_{l}$ is the
Legendre polynomial of order $l$.  In 1959, Regge~\cite{regge59, regge60} 
generalized the solution of $f$ by complexifying angular momenta.  He 
interpreted the simple poles of $a_{l}(k^{2})$ 
on the complex $l$-plane to be either resonances or bound states.

Chew and Frautschi~\cite{chew62} applied the Regge poles theory to investigate
the analyticity of $a_{l}(k^{2})$ in the case of strong interactions.
They postulated that all strongly interacting particles are self-generating
(the bootstrap hypothesis) and that they must lie on Regge trajectories 
(Chew-Frautschi conjecture)~\cite{f68}.

At first, linearity was just a convenient guide in constructing the
Chew-Frautschi plots because data were scarce and there were few
\textit{a priori} rules to direct the mesons and baryons into the same
trajectories~\cite{pC77}.  Once linearity was found to be a good
working hypothesis, justification was
given through certain assumptions in the Regge poles theory as follows:
For ${\rm Re}\; l\geq -1/2$, the partial-wave components of the scattering 
amplitude $f$ have only simple poles and
are functions of $k^{2}$,
\begin{equation}
a_{l}(k^{2}) \simeq \frac{\beta (k^{2})}{l-\alpha(k^{2})},
\end{equation}
where $\beta$ is the residue (Regge residue) and $\alpha$ the position 
(Regge trajectory) of the simple poles.
We can use Watson transformation to rewrite Eq.~[\ref{f}] as the
Sommerfeld-Watson formula~\cite{froissart} to include the poles.

A Regge trajectory $J=\alpha(k^{2})$ is also sometimes
expressed as $J=\alpha(E)$, or more commonly in terms of the
Mandelstam variable $t$ as $J=\alpha(t)$.  $t$ is 
the center-of-mass energy of the quark-antiquark pair defined
as $t\equiv (p_{q} + p_{\bar{q}})^{2}$.  It is used
instead of $s$ or $u$ because Regge poles generally arise in this
channel.  For the purpose of plotting, we use $J=\alpha(M^{2})$.  
$\alpha(t)$ represents
a set of leading Regge poles on the complex $l$-plane and is called the 
Reggeon.  The condition $\alpha(t)<0$ does not correspond to any
physical particles because $J$ cannot be negative~\cite{forshaw97}.  

Many authors~\cite{pC77, chiu72} 
are careful to disclaim linearity as being only
approximate.  For others, linearity is simply stated~\cite{segre, veseli}.  
By and large,
it is believed that Regge trajectories for relativistic scattering are 
straight lines over a considerable range of energy without any sign
of deviation~\cite{cornwall, kang, martin}.
Attempts have been made to validate this belief on 
computational grounds.  Kahana, Maung and Norbury~\cite{jN93} 
calculated the numerical solutions of the relativistic Thompson equation 
which yield linear, non-intersecting and parallel Regge trajectories.  
Their calculation did not include the effects of spin which can be a factor
in predicting the shape of the Regge trajectories.  But earlier in 1985, 
Godfrey and Isgur~\cite{godfrey85} 
solved a relativized Schr\"{o}dinger equation which did include the spin-spin 
and spin-orbit interactions.  Their calculation concurred with the results of 
linear Regge trajectories obtained from spinless particles.
These two works together seem to suggest that the effect of spin on
the shape of the Regge trajectories is negligible.  On the other hand,
if the coupling constants are not negligible, one expects the spin-orbit
contributions to be significant for high $J$ values.
Salvo~\cite{salvo94} \emph{et al.} published solutions for non-linear 
Regge trajectories by including spin dependent 
terms in a 3-dimension reduction of the Bethe-Salpeter equation.
Salvo's conclusion differs from those of Kahana and Godfrey 
concerning the effect of spin on Regge trajectories.

The linearity of Regge trajectories has been the object of investigation
once again recently.  On the theoretical front, Tang~\cite{tang93} used 
perturbative QCD to show that Regge trajectories are
non-linear by studying high-energy elastic scattering with mesonic exchange
in the case of both fixed and running coupling constants.  
On the experimental side, Brandt~\cite{UA8} \emph{et al.} affirmed the 
existence of non-linear Pomeron trajectories from
the data analysis of the recent UA8 and ISR experiments at CERN.
They published a parameterization of Pomeron trajectories containing a 
quadratic term,
\begin{equation}
\alpha(t)=1.10+0.25t+\alpha''t^{2}.
\end{equation}
where $\alpha''$ is a constant.
Recently, Burakovsky~\cite{bur98,bur99} presented a phenomenological string 
model for logarithmic and square root Regge trajectories.

In this paper, we check the claims of non-linear Regge trajectories by
plotting the most recent experimental data
published in the 1998 Review of Particle Physics (RPP)~\cite{pp98}.  
Our plots confirm the existence of non-linear trajectories.
Early Chew-Frautschi plots also show that Regge trajectories fan out.  
We refer to this
non-intersecting property as ``divergence.''  We also show that many 
trajectories intersect.
Kahana et al. numerically constructed a set of hypothetical Regge trajectories 
by using 
a fully relativistic Thompson equation.  They discovered that there are
differences in the properties of the trajectories obtained by
NRSE versus those by the Thompson equation.
We summarize the conclusions of Kahana et al. in Table 1 to illustrate these
differences.

\begin{center}
Table 1:  Comparisons of the predictions made by NRSE and
the relativistic\\
Thompson equation according to Kahana et al.~\cite{jN93}.
``Yes'' refers to a property \\
predicted by the theory and
``No'' is the prediction of the opposite property.
\vskip 10pt
\begin{tabular}{|l|c|c|}
\hline
& Non-relativistic  & Relativistic  \\
& Schr\"{o}dinger Equation & Thompson Equation \\
\hline
Linearity & No & Yes \\
Divergence & No & Yes \\
Parallelism & Yes & Yes \\
\hline
\end{tabular}
\label{summary}
\end{center}

When trajectories of different principal quantum numbers $N$ but all other
quantum numbers fixed are plotted together, they appear parallel.  We
call this property ``parallelism.''

\begin{center}
\bf 1.2 Construction of Regge Trajectories
\end{center}

The starting 
point for constructing a meson Regge trajectory is the meson assignment 
table in RPP (Table~13.2 on p.~110 of Ref.~\cite{pp98}).  We fix $I$ and 
flavor by selecting particles from a single column. 
From this column, we isolate different trajectories by fixing $N$ and 
spin-parity 
when we select particles with consecutive values of $J$.  For example, 
the $1^{1}S_{0}$, $1^{1}P_{1}$ and $1^{1}D_{2}$ states constitute an $N=1$ 
singlet trajectory; $1^{3}P_{0}$ and $1^{3}D_{1}$ the $N=1$ first triplet; 
$1^{3}P_{1}$ and $1^{3}D_{2}$ the $N=1$ second triplet; $1^{3}S_{1}$, 
$1^{3}P_{2}$, $1^{3}D_{3}$, and $1^{3}F_{4}$ the $N=1$ third triplet; 
$2^{3}S_{1}$ and  $2^{3}P_{2}$ the $N=2$ third triplet and so on.  
We use the experimental 
error instead of the width to measure the accuracy of the mass of a meson.  The
width measures the imaginary part of the complex energy while the
experimental error indicates the accuracy of the measurement of the mass at 
the resonance peak.  In case the mass of both the neutral and 
charged mesons are reported, the mass is taken to be the average of the 
three.  For example, the mass of $\pi(138)$ is taken to be the average mass of 
$\pi^{0}(135)$ 
and $\pi^{\pm}(140)$.  Similarly the error of mass is also taken to be the 
average of the 
errors of the two corresponding masses.  This scheme does not pose any serious 
ambiguity because the masses
and the errors of the neutral and charged mesons are usually quite close
(the difference in mass is usually $< 1\,\%$ and is $\sim 3.5\,\%$ in the 
worst case) and hence do not change our conclusions.  The error
of mass square, $dM^{2}$, is calculated from the mass $M$ and its error
$dM$ by the relation $dM^{2} = 2M dM$.  The end results are 13 trajectories 
containing 2 particles each, 4 containing 3 particles each and 4 containing 
4 particles each.  Single particle trajectories are omitted from the plots.
None of the $N=1$ second triplet trajectories are plotted because most of them
are single particle trajectories except the one containing $K_{1B}$ and 
$K_{2}(1820)$, where $K_{1B}$ is a nearly
equal $(45^{\circ})$ mixture of $K_{1}(1270)$ and $K_{1}(1400)$.  Since some
of these trajectories contained unconfirmed mesons, not all of them are used
in this paper.

The bold face entries in the assignment table refer to the mesons which
are confirmed by experiments.  The regular typeface entries correspond to 
those which are omitted from the summary table because of work in progress.  
For example, one of the regular typeface entries in the assignment table,
$f_{4}(2220)$, is listed as $f_{J}(2220)$ in the summary table because
$J$ may assume a value of either 2 or 4 depending on the final confirmation by
experiments.  There are other similar undetermined quantities in the meson 
data. This paper takes the conservative approach by using only the bold
face (confirmed) data contained in the RPP meson assignment table.

The baryon Regge trajectories are constructed from the RPP baryon assignment
table (Table~13.4 on p.~112 of Ref.~\cite{pp98}).  Baryons are 
categorized into 4 different
confidence levels according to their likelihood of existence.  Confidence level
1 is assigned to the baryons which are deemed the least likely to exist and
level 4 the most likely to exist.  The baryon assignment table 
contains only the level 3 and 4 particles.  These are the baryons we will 
analyze in this paper.

The baryon assignment table uses a set of slightly different quantum numbers, 
such as $J^{P}$, $(D,L^{P}_{N})$ and S.  As before, $J$ is
the total angular momentum, $P$ the parity, $L$ the orbital angular momentum
and $S$ the spin.  The new quantum number $D$ is the dimensionality of the 
symmetry group and has the value of either 56 or 70.  These numbers come from
the dimensionalities of the irreducible representations of flavor-spin $SU(6)$,
i.e. ${\bf 6}\otimes{\bf 6}\otimes{\bf 6}={\bf 56}_{S}\oplus{\bf 70}_{M}\oplus
{\bf 70}_{M}\oplus{\bf 20}_{A}$ where the subscript $S$ stands for 
``symmetric'', $A$ for ``asymmetric'' and $M$ for ``mixed symmetry.''  $N$ is 
the ``band'' which gives the number of excitation quanta.  The construction of
a baryon Regge trajectory is similar to that of the meson in that all quantum
numbers except $J$ are fixed along a trajectory.  In other words, $D$, 
$S$, flavor, strangeness
and isospin are constant along a baryon Regge trajectory.  Only $L$ is allowed 
to vary.  $N$ changes with $L$ in the same integer steps so that a change in
$N$ is the same as a change in $L$.  Hence we can ignore the consideration 
of $N$.  

Regge recurrences are separated by 2 units of $J$.  In the case of
mesons, we can plot two trajectories together in some cases because the
cross channel forces between them vanish.  It is known as the ``exchange
degeneracy'' (EXD)~\cite{chiu72} which arises out of the cross channel 
forces which split
$a(l,k)$ into even (+) and odd ($-$) signatures as $a_{\pm}(l,k)$.  
The separation of the
even and odd signatures correspond to the two different Regge trajectories.
If the cross channel forces vanish (as in the case of mesons), the even and 
odd signatures coincide and the even and odd trajectories overlap.  It means
$\alpha_{+} = \alpha_{-}$ and $\beta_{+} = \beta_{-}$.  These are called the
EXD conditions.  When the EXD conditions apply, the even and odd parity
mesons can be plotted along the same trajectories.  In the case of baryons,
the cross channel forces persist.  Therefore the even and odd parity baryons
cannot be plotted together in the same trajectories. 
The EXD criteria enable us to pick out 3 trajectories of 3 baryons each and 2
trajectories of 2 baryons each from the baryon assignment table.  These
selections are achieved by isolating a column (e.g. the $N(939)$--$N(2220)$
column)
and picking all the particles with the same $D$, $S$ and $P$ (e.g. $N(939)$,
$N(1680)$ and $N(2220)$).  Once a trajectory is picked from the first column,
corresponding entries of the following columns also constitute baryon Regge 
trajectories.  The spectroscopic notation for baryons is $L_{2I,2J}$.
The $N$ and $\Lambda$ trajectories are made up of the $P_{11}$, $F_{15}$,
$H_{19}$ states and the $\Delta$ trajectory is made up of 
the $P_{33}$, $F_{37}$, $H_{3\,11}$ states. 

\section{Linearity}

Linearity means that all the
particles of a Regge trajectory must lie on the straight line
$M^{2}=\alpha J + \beta$.  In graphical
analysis, non-linearity can be detected by simple inspection in only
extreme cases.  Linearity on the other hand is more difficult to judge.  
Therefore we
devise a method called the ``zone test'' to facilitate this judgment.

\begin{center}
\bf 2.1 Zone Test
\end{center}

\begin{figure}[ht]
\begin{center}
\epsfig{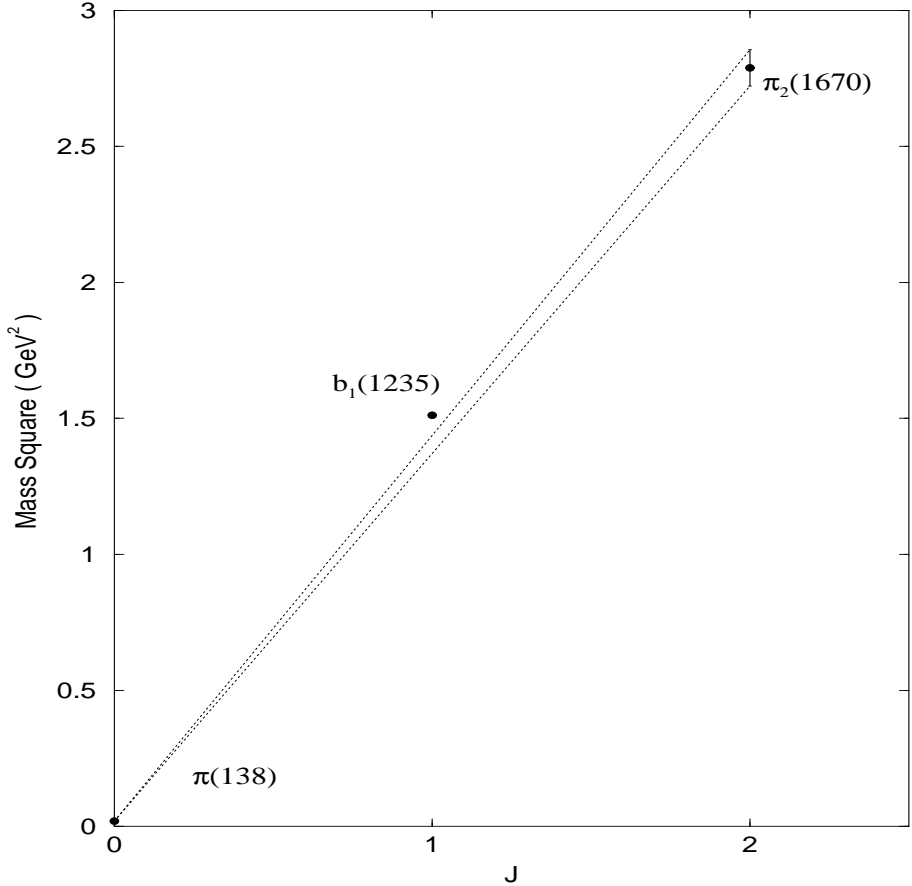}
\caption{\label{mlpi}
Meson Regge trajectory of the $N=1$, $S=0$ singlet states.  $b_{1}(1235)$ 
fails to intersect the zone.  The polynomial fit is $M^{2}=-0.1077J^{2}+1.6003J
+0.019\,(\rm GeV^{2})$.  The mass of $\pi(138)$ is taken to be the average of 
the masses of $\pi^{0}(135)$ and $\pi^{\pm}(140)$.  Although there is a 6.5~\%
difference between the mass squares of $\pi^{0}(135)$ and $\pi^{\pm}(140)$,
the test zone is virtually unchanged by this small difference because of the
large mass squared of the other two mesons on the trajectory.
The zone test suggests that the $\pi$ trajectory is non-linear.}
\end{center}
\end{figure}

\begin{figure}[ht]
\begin{center}
\epsfig{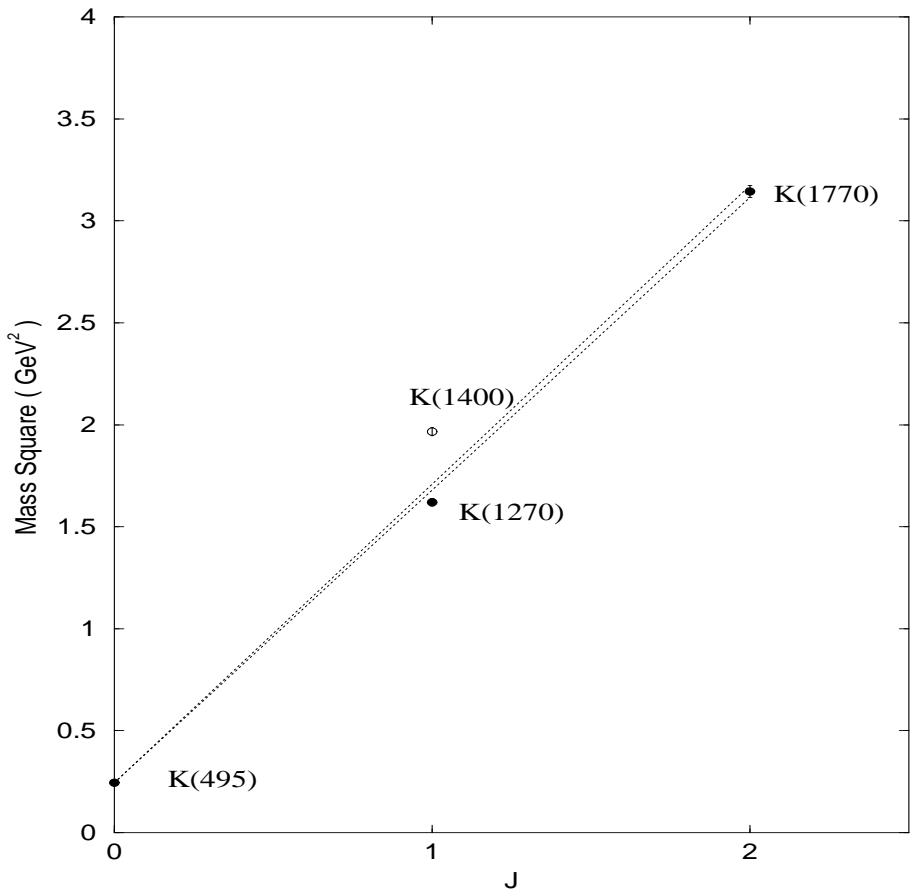}
\caption{\label{mlk}
Meson Regge trajectory of the $N=1$, $S=0$ singlet states.  The RPP assignment 
table lists $K_{1B}$ as a nearly
equal mixture of $K(1270)$ and $K(1400)$.  In the graph above, we plot
both constituent mesons at $J=1$ for the sake of completeness.
Neither $K(1270)$, $K(1400)$ nor their average satisfies the zone test by
simple inspection.  The polynomial fit of $K(495)$, $K(1270)$ and $K(1770)$
is $M^{2}=0.0737J^{2}+1.3018J+0.245\,(\rm GeV^{2})$.}
\end{center}
\end{figure}

\begin{figure}[ht]
\begin{center}
\epsfig{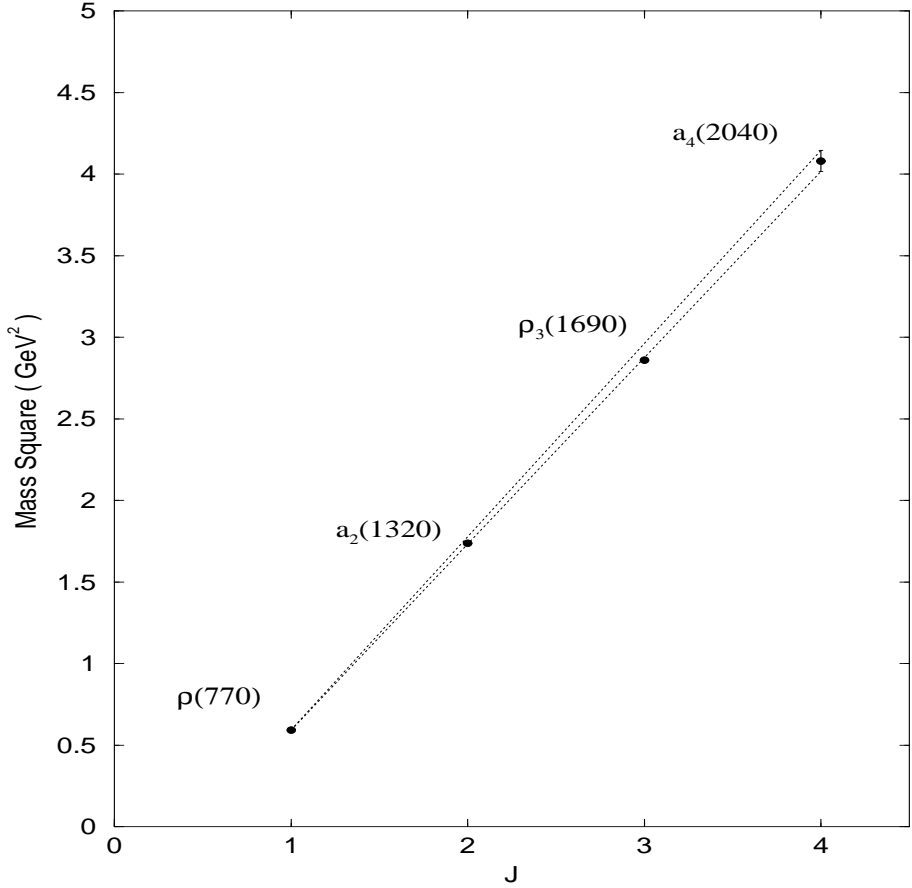}
\caption{\label{mlrho}
Meson Regge trajectory of the $N=1$, $S=1$ third triplet states.  Both 
$a_{2}(1320)$ and $\rho_{3}(1690)$ intersect the zone and hence this
trajectory passes the zone test.  Linearity is supported by the numerical 
zone test.  The polynomial fit is $M^{2}=0.0191J^{2}+
1.0629J-0.4831\,(\rm GeV^{2})$.  The negative vertical intercept corresponds
to a non-sense pole because $J<1$ is not allowed in an $S=1$ state.}
\end{center}
\end{figure}

\begin{figure}[ht]
\begin{center}
\epsfig{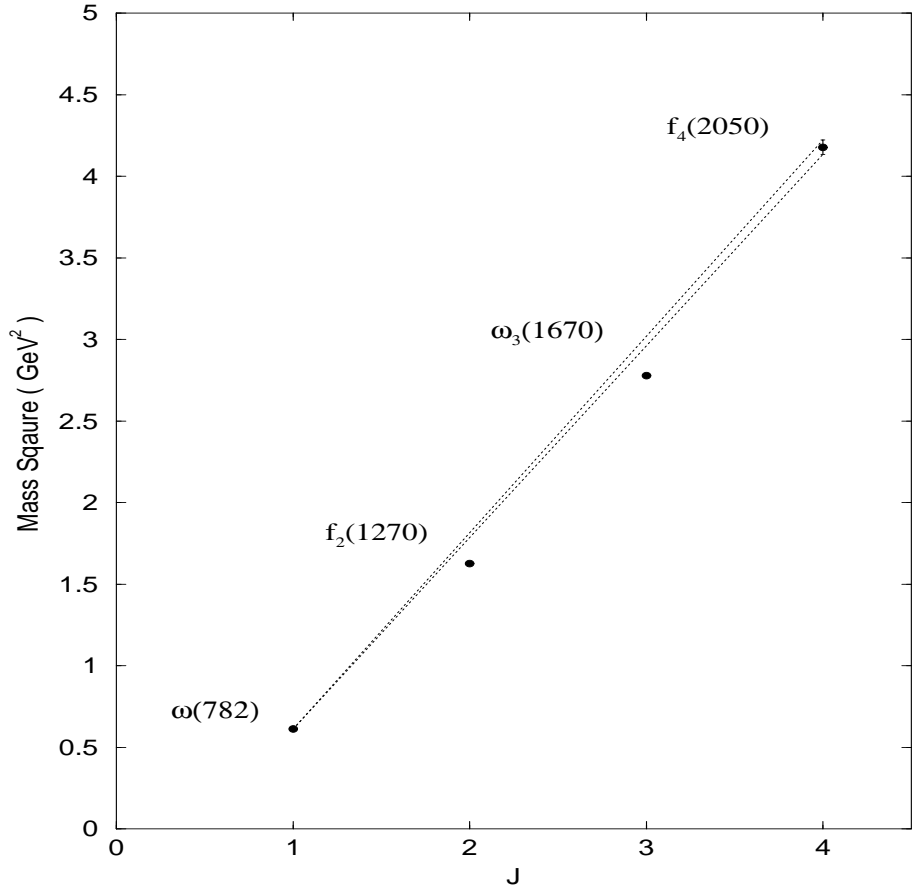}
\caption{\label{mlomega}
Meson Regge trajectory of the $N=1$, $S=1$ third triplet states.  Both 
$f_{2}(1270)$ and $\omega_{3}(1670)$ fail to intersect the zone.  The
polynomial fit is $M^{2}=0.0962J^{2}+0.7042J-0.1837\,(\rm GeV^{2})$.}
\end{center}
\end{figure}

\begin{figure}[ht]
\begin{center}
\epsfig{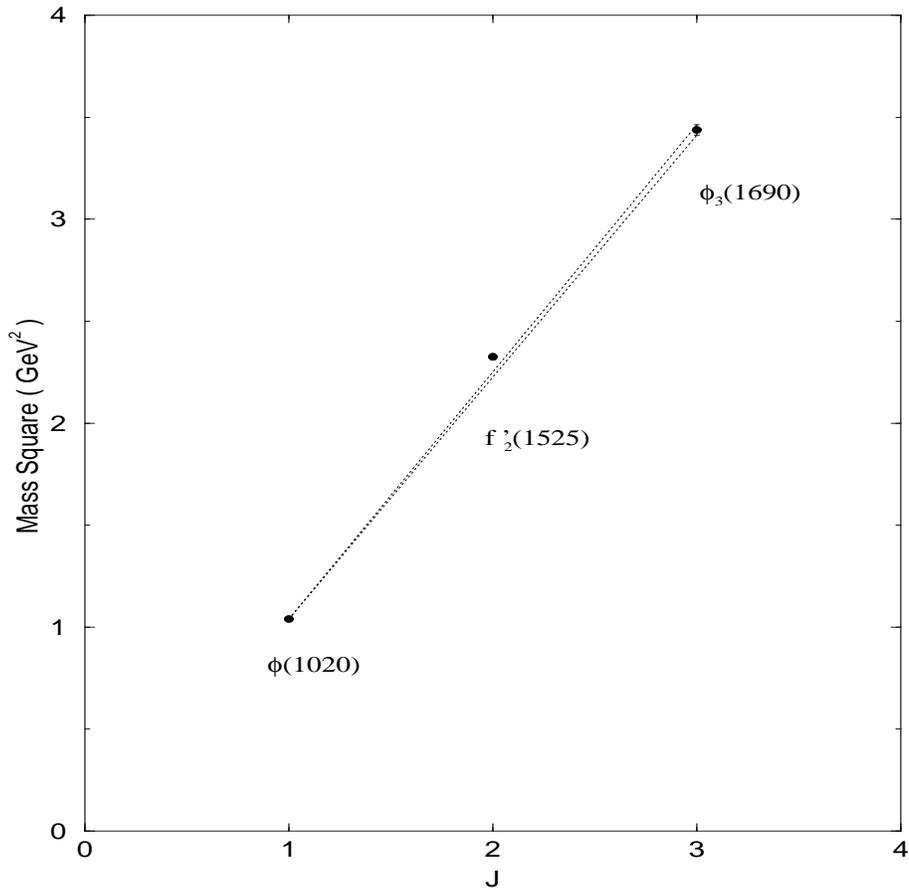}
\caption{\label{mlphi}
Meson Regge trajectory of the $N=1$, $S=1$ third triplet states.  
$f'_{2}(1525)$ fails to intersect the zone.  The
polynomial fit is $M^{2}=-0.0879J^{2}+1.5505J-0.4234\,(\rm GeV^{2})$.}
\end{center}
\end{figure}

\begin{figure}[ht]
\begin{center}
\epsfig{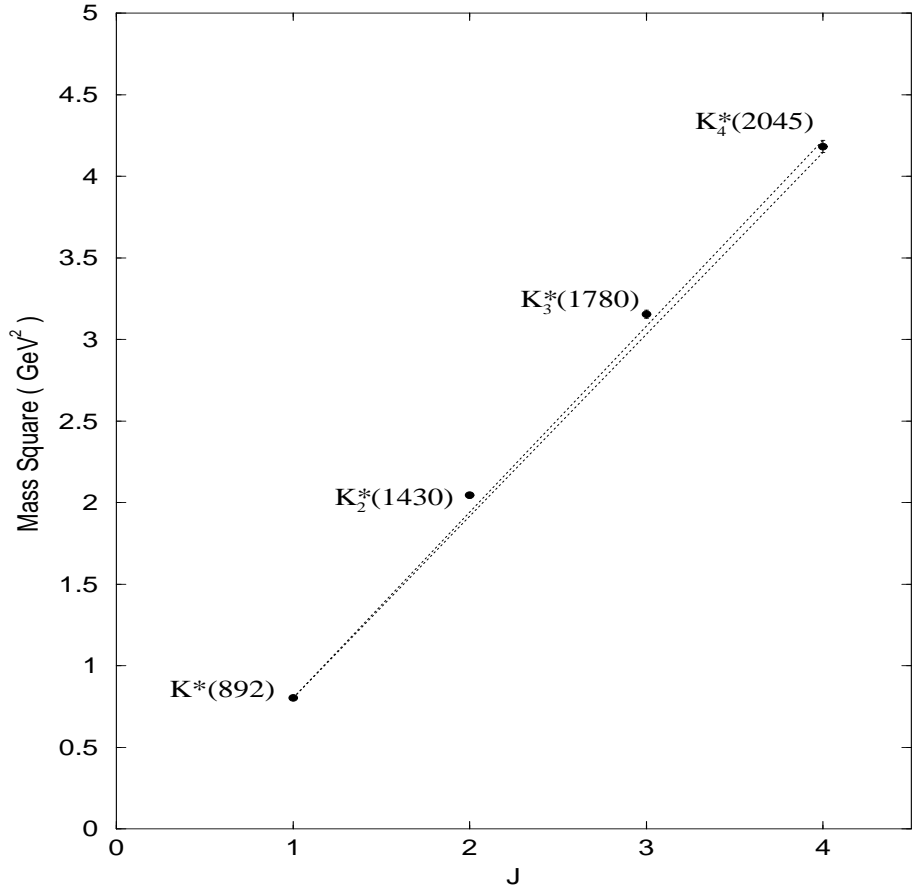}
\caption{\label{mlkstar}
Meson Regge trajectory of the $N=1$, $S=1$ third triplet states.  Both 
$K_{2}^{\ast}(1430)$ and $K_{3}^{\ast}(1780)$ fail to intersect the zone.
The polynomial fit is $M^{2}=-0.0535J^{2}+1.3922J-0.5331\,(\rm GeV^{2})$.}
\end{center}
\end{figure}

We test linearity by the ``zone test'' on Regge trajectories with 3 or more 
particles.  A test zone of an experimental Regge trajectory is defined to be 
the area bounded by the 
error bars of the first and the last particles and the straight lines joining 
them.  Figs.~\ref{mlpi}--\ref{mlkstar} illustrate these test zones (regions
enclosed by the dotted lines).  
A zone contains all the possible 
straight lines crossing the error bars of the first and the last particles.  A
Regge trajectory can be a straight line if the error bars of all 
other particles intersect the zone.  In most cases, intersections are easily 
discernible by inspection.  If ambiguity
ever arises in borderline cases, an exact numerical version of the zone test is
used.  

Suppose we are given a sequence of $N$ mesons and their values of mass square 
with errors, $\{ M_{i}^{2} \pm dM_{i}^{2} \}$.  We calculate the 
equation of the straight line connecting $M_{1}^{2} + dM_{1}^{2}$ and
$M_{N}^{2} + dM_{N}^{2}$ and then the equation of the line connecting
$M_{1}^{2} - dM_{1}^{2}$ and $M_{N}^{2} - dM_{N}^{2}$.  These two lines
define the boundaries of the zone.  For each $J$, we can calculate the bounds
to be intersected by the error bar to qualify as a linear Regge trajectory.  
For 
a 3-particle trajectory in which the particles are labelled $(1,2,3)$, the 
lower and upper bounds at $J=2$ are calculated as
\begin{eqnarray}
lb(3,2) &=& {{(M_{1}^{2} - dM_{1}^{2}) + (M_{3}^{2} - dM_{3}^{2})} \over 2 }, 
\nonumber \\
ub(3,2) &=& {{(M_{1}^{2} + dM_{1}^{2}) + (M_{3}^{2} + dM_{3}^{2})} \over 2 },
\end{eqnarray}
where $lb(3,2)$ stands for the lower bound and $ub(3,2)$ the upper bound of 
particle 2 along a 3-particle trajectory.  Similarly, we can calculate the 
bounds of particles 2 and 3 along a 4-particle trajectory as
\begin{eqnarray}
lb(4,2) &=& {{2(M_{1}^{2} - dM_{1}^{2}) + (M_{4}^{2} - dM_{4}^{2})} \over 3 }, 
\nonumber \\
ub(4,2) &=& {{2(M_{1}^{2} + dM_{1}^{2})+(M_{4}^{2} + dM_{4}^{2})} \over 3 }.\\
lb(4,3) &=& {{(M_{1}^{2} - dM_{1}^{2})+2(M_{4}^{2} - dM_{4}^{2})} \over 3 }, 
\nonumber \\
ub(4,3) &=& {{(M_{1}^{2} + dM_{1}^{2}) + 2(M_{4}^{2} + dM_{4}^{2})} \over 3 }.
\end{eqnarray}
We can generalize these results for particle $i$ along an $n$-particle 
trajectory as
\begin{eqnarray}
lb(N,i) &=& {(N-i)(M_{1}^{2}-dM_{1}^{2})+(i-1)(M_{N}^{2}-dM_{N}^{2})} 
\over {N-1}, \nonumber \\
ub(N,i) &=& {(N-i)(M_{1}^{2}+dM_{1}^{2})+(i-1)(M_{N}^{2}+dM_{N}^{2})} 
\over {N-1}.
\end{eqnarray}
This numerical method is useful for checking linearity when simple inspection
is inconclusive.

\begin{center}
\bf 2.2 Conclusions from Zone Test
\end{center}

All of the data points in all of the graphs in
this paper are shown with error bars.  If the error bars are invisible in the
plots, it simply means that the error bars are smaller than the symbols
of the associated data points.  We use the zone test to check linearity by 
simple inspection in Figs. \ref{mlpi}--\ref{mlkstar}.  
At least one of the error bars of the intermediate particles fails
to intersect the test zone in all of the figures except Fig. \ref{mlrho}.

Figs. \ref{mlpi}--\ref{mlk}
illustrate a group of meson Regge trajectories of the $N=1$, $S=0$ 
singlet states and varying $J$ corresponding to the $1^{1}S_{0}$, $1^{1}P_{1}$ 
and $1^{1}D_{2}$ states.  Both trajectories fail the zone test and are
non-linear.  The $\pi$ trajectory has a decreasing slope.

In Figs. \ref{mlrho}--\ref{mlkstar}, 
trajectories of the $N=1$, $S=1$ third triplet states with
varying $J$ corresponding to the $1^{3}S_{1}$, $1^{3}P_{2}$, $1^{3}D_{3}$, 
and $1^{3}F_{4}$ states are plotted.  Trajectories in Figs. \ref{mlomega}
and \ref{mlkstar} fail the zone test by simple inspection.  In Fig. 
\ref{mlrho},
the error bars of both $a_{2}(1320)$ and $\rho_{3}(1690)$ appear to intersect 
the zone at the lower boundary.  In this case, the numerical version of the 
zone test is used.  The error bars of both particles
must intersect the bounds to support linearity. The bounds 
in the case of $a_{2}(1320)$ are $(lb,ub)(4,2)=(1.73, 1.78)\rm\, GeV^{2}$ 
which intersect the error bar, (1.7358, 1.7390)$\rm\, GeV^{2}$.  The bounds for
$\rho_{3}(1690)$ are $(lb,ub)(4,3)=( 2.87, 2.96 )\rm\, GeV^{2}$ which 
also intersect the error bar, (2.843, 2.876)$\rm\, GeV^{2}$.  The
numerical test supports the existence of a straight line intersecting 
all the error bars of the particles along this trajectory.  The $\omega$
trajectory has an increasing slope while both the $\phi$ and $K^{*}$
trajectories have decreasing slopes.

\begin{figure}[ht]
\begin{center}
\epsfig{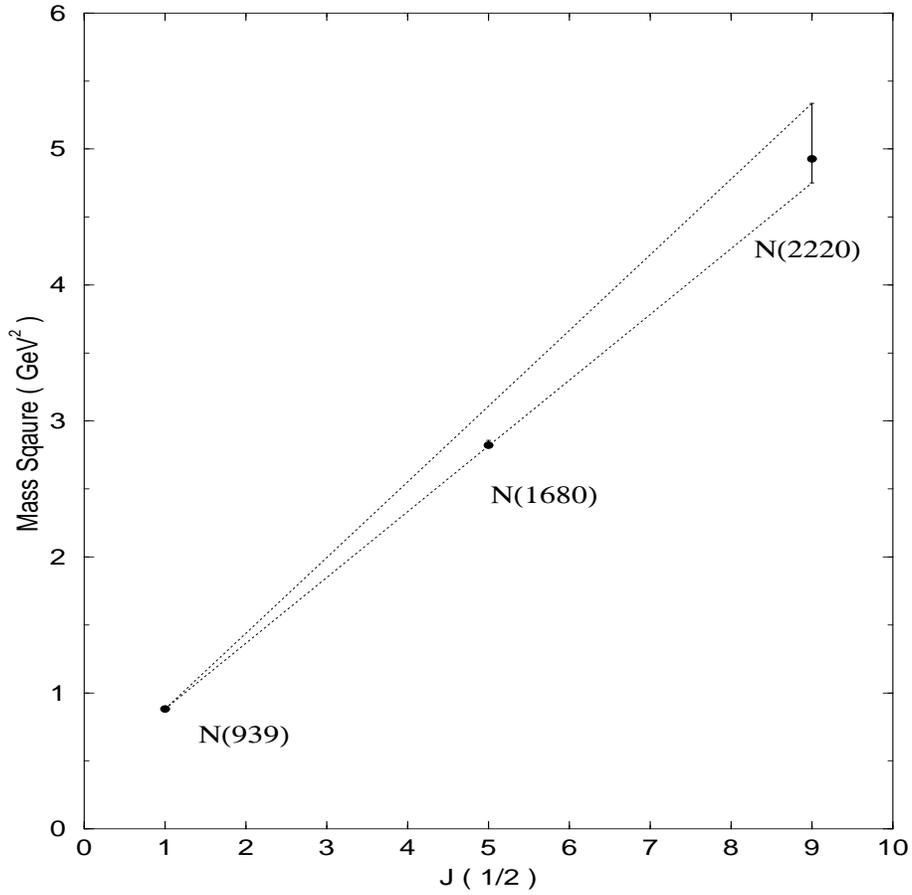}
\caption{\label{bln1}
Baryon Regge trajectory of $S=1/2$, $P=+$ and $I=1/2$ octet states.  
$N(1680)$ satisfies the zone test by simple 
inspection.  The polynomial fit is $M^{2}=0.0207J^{2}+0.9081J+0.4223$ (GeV).}
\end{center}
\end{figure}

\begin{figure}[ht]
\begin{center}
\epsfig{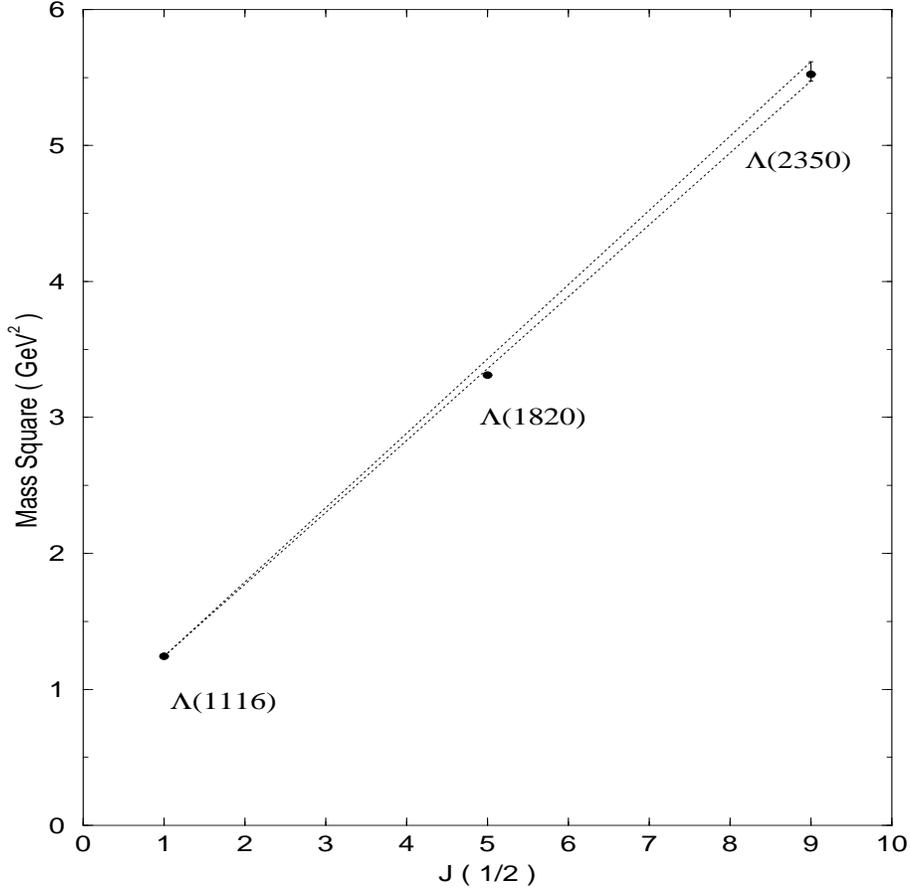}
\caption{\label{bllambda1}
Baryon Regge trajectory of $S=1/2$, $P=+$ and $I=0$ octet states.  
$\Lambda(1820)$ has a numerical bound of 
$(lb,ub)(3,2)=(3.360,3.431)\,\rm GeV^{2}$ which does not intersect the 
experimental bound of $(3.294,3.331)\,\rm GeV^{2}$.
$\Lambda(1820)$ fails the numerical zone test.   The
polynomial fit is $M^{2}=0.0180J^{2}+0.9797J+0.7524$ (GeV).}
\end{center}
\end{figure}

The zone test for baryon trajectories are illustrated in 
Figs.~\ref{bln1}--\ref{bldelta1}.  The $N$ and $\Delta$ trajectories in 
Figs.~\ref{bln1} and \ref{bldelta1} clearly satisfy the zone test by simple 
inspection.  The $\Lambda$
trajectory in Fig.~\ref{bllambda1} is shown to be non-linear by the numerical 
zone test.  In summary, 6 of 8 trajectories with 3 or more particles each
are shown to be non-linear.  Polynominal fits of the trajectories are
included in the figure captions for reference only.

\section{Divergence}

Divergence seems to be a property of the Regge trajectories in the early
Chew-Frautschi plots and is also a prediction of the numerical calculations
by Kahana et al~\cite{jN93}.  Divergence is defined to be the conjunction of 
two properties: (1)~non-intersection and (2)~fanning out.

\begin{figure}[ht]
\begin{center}
\epsfig{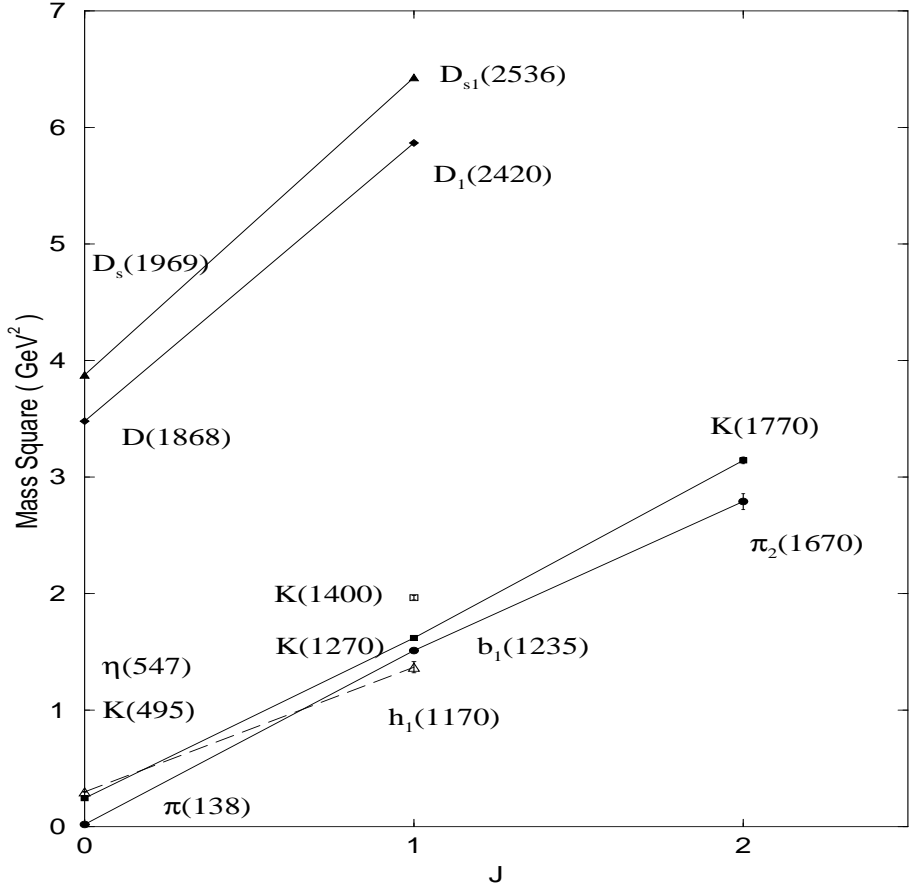}
\caption{\label{md1}
Meson Regge trajectories of the $N=1$, $S=0$ singlet states.  The series
consists of the $1^{1}S_{0}$, $1^{1}P_{1}$ and $1^{1}D_{2}$ states.  The 
trajectories include those illustrated in Figs. \ref{mlpi} and \ref{mlk}
as well as the $D$ mesons.  The $D$ and $D_{s}$ mesons form a group
and the $K$ and light unflavored mesons $\eta$ and $\pi$ form another.
Global divergence is observed among groups of trajectories of large mass
difference but local divergence is violated when the $\eta$ trajectory 
(denoted by the dotted line) intersects the $K$ and $\pi$ trajectories.}
\end{center}
\end{figure}

\begin{figure}[ht]
\begin{center}
\epsfig{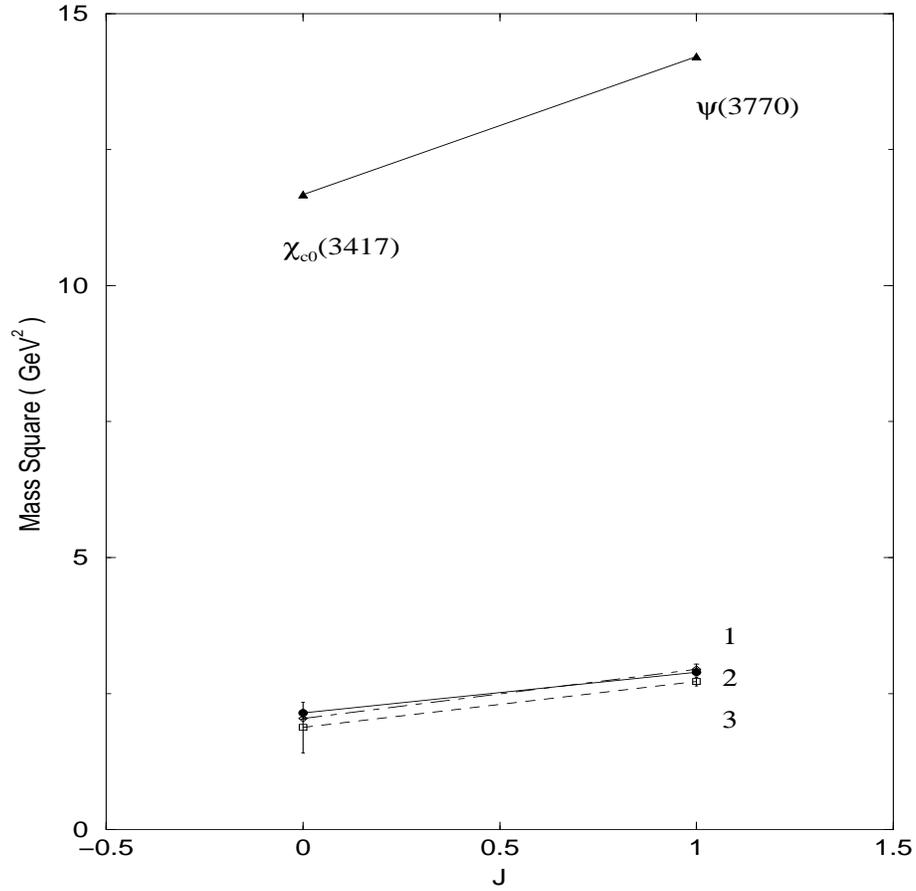}
\caption{\label{md2a}
Meson Regge trajectories of the $N=1$, $S=1$ first triplet states.  The 
series consists of the $1^{3}P_{0}$ and $1^{3}D_{1}$ states.  The
trajectories labelled 1--3 are magnified in Fig. \ref{md2b}.
Global divergence is observed in this graph.}
\end{center}
\end{figure}

\begin{figure}[ht]
\begin{center}
\epsfig{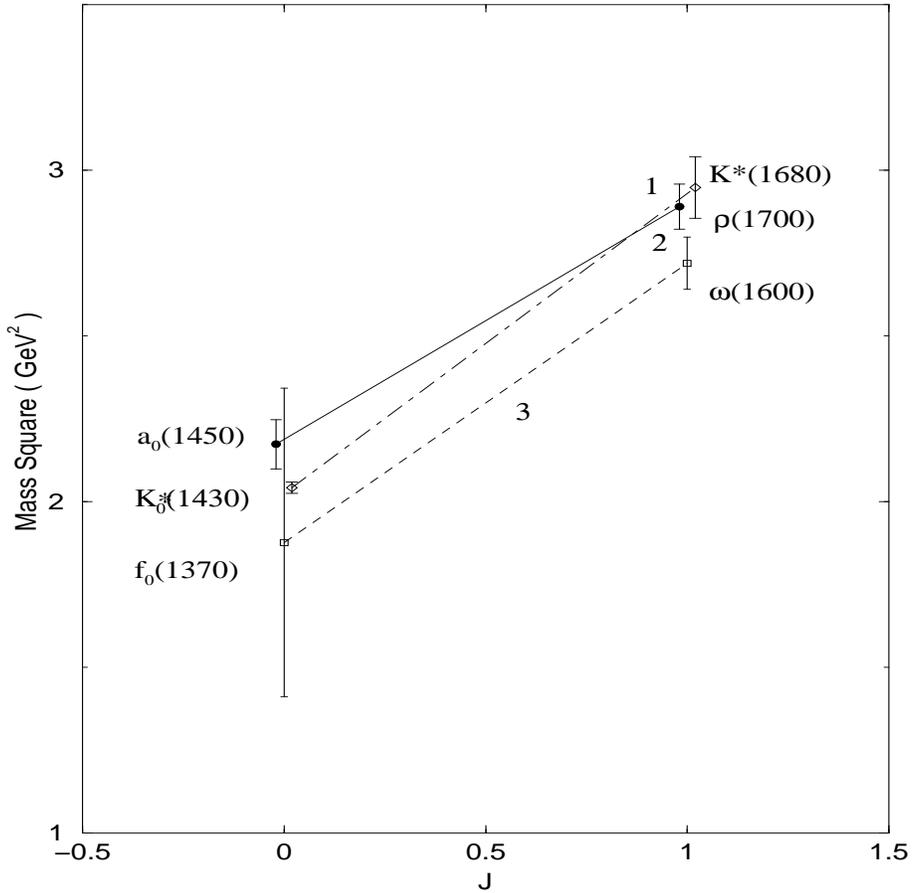}
\caption{\label{md2b}
Meson Regge trajectories of the $N=1$, $S=1$ first triplet states.  The 
series consists of the $1^{3}P_{0}$ and $1^{3}D_{1}$ states.  The trajectories
1--3 are the magnifications of a subset of Fig.~\ref{md2a}.  
Trajectories 1 and 2 are
shifted horizontally slightly to separate the error bars.
Divergence is inconclusive because of the large error bars.  The actual mass 
of $K^{\ast}(1680)$ is 1717 MeV which causes it to 
appear higher than $\rho(1700)$ in the graph.}
\end{center}
\end{figure}

\begin{figure}[ht]
\begin{center}
\epsfig{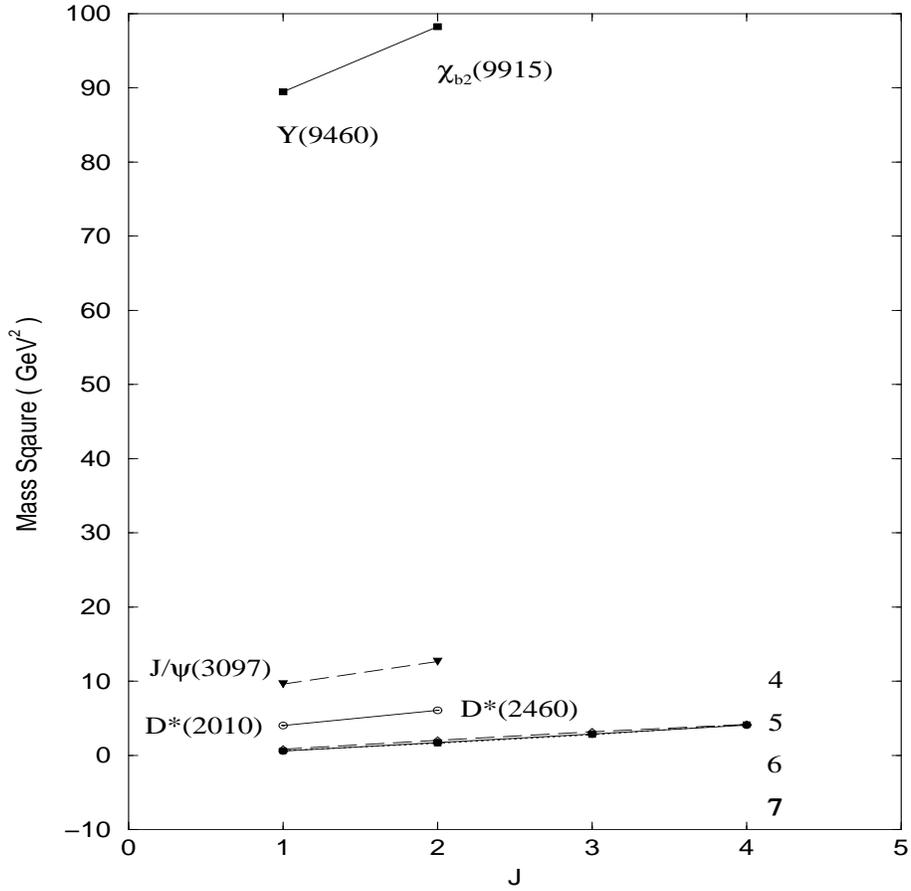}
\caption{\label{md3a}
Meson Regge trajectories of the $N=1$, $S=1$ third triplet states.  The 
series consists of the $1^{3}S_{1}$, $1^{3}P_{2}$, $1^{3}D_{3}$, and 
$1^{3}F_{4}$ states.  The group of trajectories
labelled 4--7 are magnified in Fig. \ref{md3b}.  Global divergence
is observed.}
\end{center}
\end{figure}

\begin{figure}[ht]
\begin{center}
\epsfig{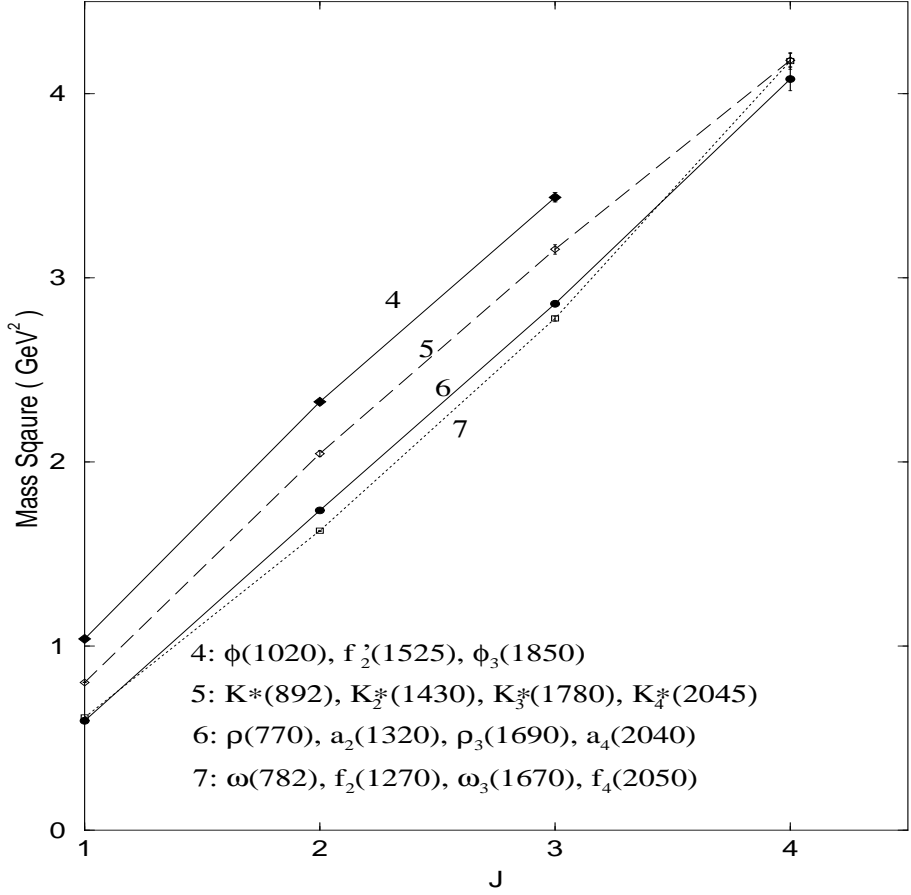}
\caption{\label{md3b}
Meson Regge trajectories of the $N=1$, $S=1$ third triplet states.  The 
series consists of the $1^{3}S_{1}$, $1^{3}P_{2}$, $1^{3}D_{3}$, and 
$1^{3}F_{4}$ states.  The trajectories 4--7 are magnifications of a subset of 
the trajectories in Fig. \ref{md3a} and are the
same trajectories as in Figs. \ref{mlrho}--\ref{mlkstar}.
Divergence is violated by these trajectories.}
\end{center}
\end{figure}

\begin{figure}[ht]
\begin{center}
\epsfig{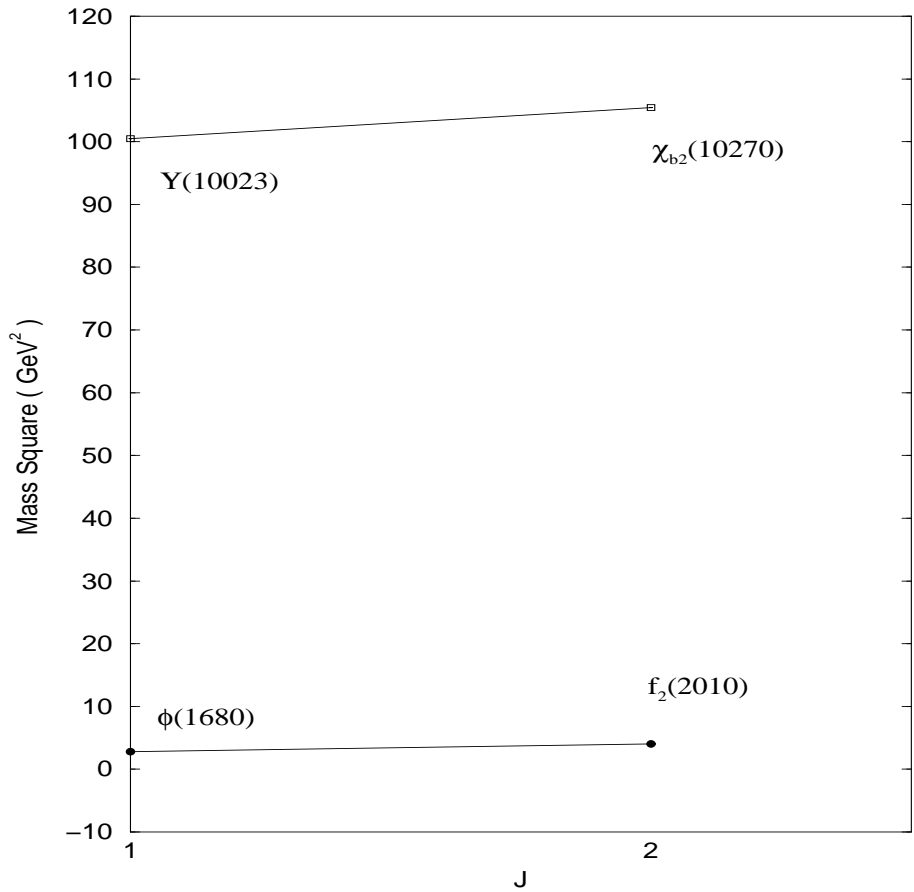}
\caption{\label{md4}
Meson Regge trajectories of the $N=2$, $S=1$ third triplet states.  The 
series consists of the $2^{3}S_{1}$ and $2^{3}P_{2}$ states.  
Divergence is observed.}
\end{center}
\end{figure}

We check divergence by plotting families of meson Regge
trajectories with the same isospin and spin-parity in Figs. 
\ref{md1}--\ref{md4}.  It is observed that
non-linear trajectories of similar masses intertwine.  
In general, Regge trajectories are not evenly saparated in a graph.  Some
trajectories can be obscured when many of them are plotted over a large mass
range on the same graph.  We adopt a numeration scheme which allows us 
to identify the obscured trajectories in
separate plots.  For example, the group denoted as 1--3 in Fig. \ref{md2a} 
is magnified as trajectories 1--3 in Fig. \ref{md2b}.
Divergence is clearly violated in Fig. \ref{md3b} when
trajectories intersect.  Due to the large error bars, divergence
in Fig. \ref{md2b}, the determination of the properties
of these meson trajectories is inconclusive.

\begin{figure}[ht]
\begin{center}
\epsfig{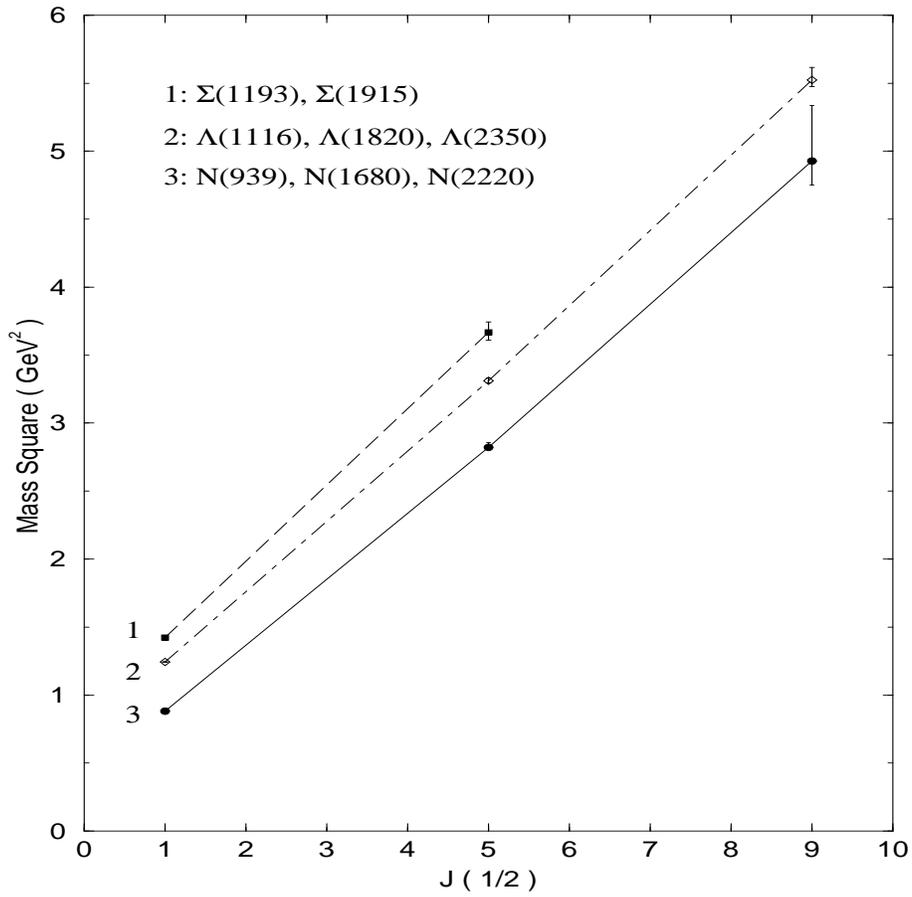}
\caption{\label{bd1}
Baryon Regge trajectories of $P=+$ octet states.  Due to the scarcity of 
data and the large error of $N(2220)$, divergence is plausible but 
inconclusive.}
\end{center}
\end{figure}

\begin{figure}[ht]
\begin{center}
\epsfig{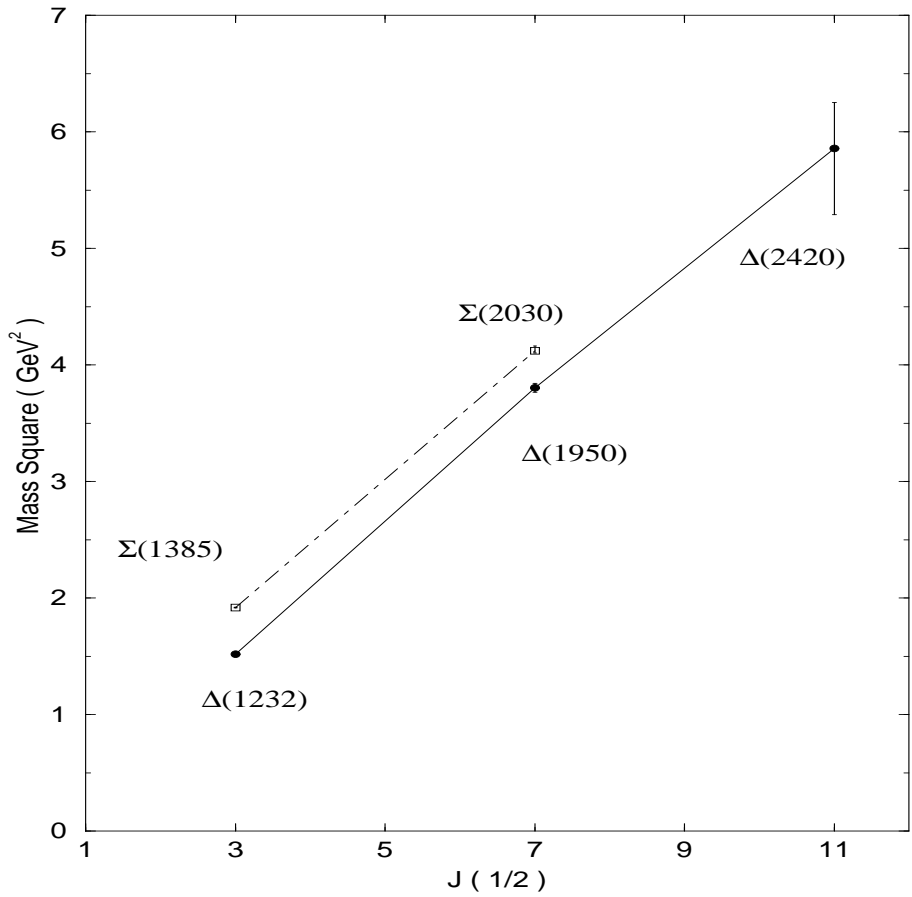}
\caption{\label{bd2}
Baryon Regge trajectories of $P=+$ decuplet states.  Due to the scarcity of 
data and the large error of $\Delta(2420)$, divergence is plausible but 
inconclusive.}
\end{center}
\end{figure}

Although individual meson trajectories do not fan out, it can be seen in
Figs. \ref{md1}, \ref{md2a}, \ref{md3a} and \ref{md4} that groups of
them diverge on a global level.  We also notice that these groups can be 
labelled according to mass difference.  In general, the mass of the
intersecting trajectories does not differ significantly.  On the other
hand, divergent trajectories have large mass difference.    
For example, in Fig.~\ref{md1}, the $\pi$, $K$ and 
$\eta$ trajectories have small mass difference and form a group of
intersecting trajectories.  The $D$ and $D_{S}$ trajectories also form a 
group with small mass difference.  These two groups of trajectories diverge
globally.  In summary, trajectories of small mass difference do not diverge
but those of large mass difference fan out in our plots.
In the case of baryon Regge trajectories, there are insufficient data to
test divergence.  Divergence is shown to be plausible
in Figs.~\ref{bd1} and \ref{bd2}.

\section{Parallelism}

\begin{figure}[ht]
\begin{center}
\epsfig{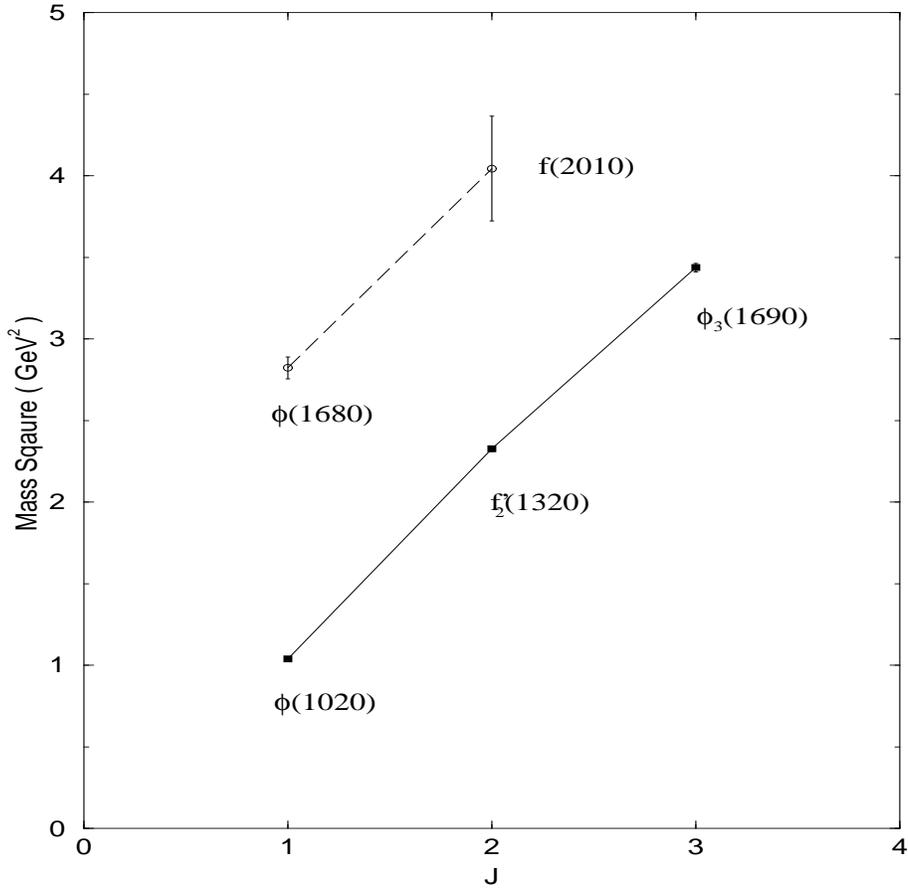}
\caption{\label{mp}
Meson Regge trajectories of the $N=1,2$, $S=1$ third triplet states.  The 
series consists of the $1^{3}S_{1}$, $1^{3}P_{2}$, $2^{3}S_{1}$ and 
$2^{3}P_{2}$ states.  The $N=1$ states are denoted by solid lines and
$N=2$ states by long-dashed lines.  Parallelism is inconclusive
due to the large error bar of $f(2010)$.}
\end{center}
\end{figure}

Parallelism refers to the property that Regge trajectories of different
values of $N$ (which are otherwise identical) are parallel.  Two
trajectories are parallel if the dynamics are similar.  
There is no \textit{a priori} reason why 
parallelism must hold.  There are only
two $\phi$ trajectories with $N=1$ and $N=2$ which qualify for this test.  
Fig.~\ref{mp} shows that the two trajectories appear to be parallel.  
However these trajectories
consist of only 2 or 3 mesons each.  It is not clear how they will behave
at $J>2$.  The error of $f_{2}(2010)$ is also quite large compared to the
separation of the two trajectories.  In conclusion, the status of
parallelism as a candidate for a property of Regge trajectories is still
uncertain.

\section{Conclusion}

The linearity of Regge trajectories is clearly violated in Figs. \ref{mlpi},
\ref{mlk}, \ref{mlomega} and \ref{mlkstar} by simple inspection but is 
supported by the numerical zone test
in Fig. \ref{mlrho}.  Divergence is not observed on an individual basis.  
On the other hand, divergence of groups of trajectories of small mass
difference is observed on a global level.  Due to insufficient data, 
parallelism is inconclusive.

Currently there are a variety of models predicting both linear and
non-linear Regge trajectories.  In general, almost all theories~\cite{salvo94,
andreev99,bur99,grosse97,filipponi98,prosperi99,baldicchi99} 
agree that meson Regge trajectories are linear in the small $J$ limit.  
Our plots contradict these predictions.  Secondly, all non-linear Regge 
trajectories models predict trajectories with either increasing or decreasing 
slopes exclusively, but not both~\cite{salvo94,bur99,inopin99}.  Our plots 
show that meson Regge trajectories of both kinds exist.  Therefore data
rule out all the models that predict non-linear meson Regge trajectories with 
strictly increasing or decreasing slopes.  In the end, data rule
out all current meson Regge trajectories models because they are faced with
at least one of the problems mentioned above.

\section{Acknowledgment}
We thank Dr. Sudha Swaminathan and Prof. Dale Snider for their comments.
This work was supported in part by NASA Research Grant Numbers NCC-1-354
and NCC-1-260.

\newpage

\newpage

\newpage

\newpage

\newpage

\newpage

\newpage

\newpage

\newpage

\begin{figure}[ht]
\begin{center}
\epsfig{file=bldelta1.pstex,width=12cm,height=12cm}
\caption{\label{bldelta1}
Baryon Regge trajectory of $S=3/2$, $P=+$ and $I=3/2$ decuplet states.  
$\Delta(1950)$
satisfies the zone test due to the large error of $\Delta(2420)$.  
The polynomial fit is $M^{2}=-0.029J^{2}+1.2875J-0.3480$ (GeV).}
\end{center}
\end{figure}

\newpage

\newpage

\newpage

\newpage

\newpage

\newpage

\newpage

\newpage

\newpage

\end{document}